# An Networked HIL Simulation System for Modeling Large-scale Power Systems

Fuhong Xie, Catie McEntee, Mingzhi Zhang, Ning Lu
Electrical & Computer Engineering Department
North Carolina State University
Raleigh, North Carolina, USA
fxie2@ncsu.edu, nlu2@ncsu.edu

Xinda Ke, Mallikarjuna R. Vallem, Nader Samaan
Electricity Infrastructure
Pacific Northwest National Laboratory
Richland, Washington, USA
xke@pnnl.gov, nader.samaan@pnnl.gov

*Abstract* — **This paper presents a network hardware-in-the-loop (HIL) simulation system for modeling large-scale power systems. Researchers have developed many HIL test systems for power systems in recent years. Those test systems can model both microsecond-level dynamic responses of power electronic systems and millisecond-level transients of transmission and distribution grids. By integrating individual HIL test systems into a network of HIL test systems, we can create large-scale power grid digital twins with flexible structures at required modeling resolution that fits for a wide range of system operating conditions. This will not only significantly reduce the need for field tests when developing new technologies but also greatly shorten the model development cycle. In this paper, we present a networked OPAL-RT based HIL test system for developing transmission-distribution coordinative Volt-VAR regulation technologies as an example to illustrate system setups, communication requirements among different HIL simulation systems, and system connection mechanisms. Impacts of communication delays, information exchange cycles, and computing delays are illustrated. Simulation results show that the performance of a networked HIL test system is satisfactory.**

## I. INTRODUCTION

THE increasing penetration of distributed energy resources (DERs) makes power grid operation more flexible, but at the same time more complex. To access the control flexibility of DERs from different levels, it is critical to aggregate them from the single device level up to the transmission system level. Therefore, co-simulation of transmission systems, distribution grids, and DER systems becomes critical in power system planning and operation studies. Moreover, conducting tests and experiments in testing facilities for developing and testing new power system technologies is not only prohibitively expensive but also risky because of the lack of experiences and possible design flaws.

Therefore, in recent years, many research efforts have been devoted to the development of high fidelity, real-time hardware-in-the-loop (HIL) test systems for developing and testing new controllers and studying the interactions between DERs and power systems. This approach is attractive because a test system developed on a HIL-based platform allows actual controllers and devices to be tested with appropriate communication links in real-time. Thornton M et al. [1] used internet-of-things and HIL simulation to construct a load node to represent part of the simulated system. Kim Y J et al. [2] developed a power HIL testbed to analyze the effects of direct load control on real-time grid frequency regulation, where actual heat pumps and energy storage systems are used to mitigate load variations.

However, using a standalone HIL testbed for modeling large-scale complex systems has some drawbacks. *First*, the cost for setting up a standalone multi-core OPAL-RT HIL simulator typically is more than $100,000, making it prohibitively expensive for a research group to own many multi-core HIL simulators for scaling up the simulation to model large-scale, complex control systems. For example, the computational power required to co-simulate both the transmission system and the distribution system with high DER penetration increases linearly with the number of systems to be modeled. This usually can be mitigated only by increasing the number of the simulation core. *Second*, researchers have distinct research focuses. For example, power system engineers are interested in modeling the system level response of power systems while power electronics engineers are interested in designing controls and circuits for converter and inverter.

To model a large-scale power system that contains many power electronic systems, it is economical and highly scalable to establish a networked HIL simulation system that only represents the selected subsystem in detail. This networked HIL test system should allow asynchronous simulation steps in different systems, model interactions between systems, and allow information exchange across systems.

Thus, in this paper, we propose a framework for setting up a networked HIL test system for simulating large-scale power systems. The setup of a coordinated real-time sub-transmission Volt-VAR control (VVC) testbed (CReS-VCT) that co-simulates transmission-distribution-DER systems is used as an example to illustrate the proposed framework. Implementations and simulation models were developed by research teams at Pacific Northwest National Laboratory (PNNL), North Carolina State University (NC State), and the University of Texas at Austin (UT-Austin). Measurements and control actions among HIL simulators are communicated via a virtual private network (VPN) tunnel or a shared-file method. The contribution of this paper is the design of the coordination mechanism for setting up a networked HIL test system. This paper extends our research work in [3] by separating the whole system to different simulators and reports the impacts of communication delays, simulation time step requirements, information exchange frequency, and control sequence design on the performance of a networked HIL test system.

The rest of the paper is organized as follows. Section II provides an overview of the testbed. Section III discusses the design of the proposed testbed and the modeling of the main components. Section IV presents the case studies and analysis. Conclusions and future work are summarized in Section V.

## II. OVERVIEW FOR A NETWORK OF HIL TEST SYSTEMS

In this paper, we will use the setup of the CReS-VCT to illustrate the architecture of a networked HIL test system.

### A. HIL Test System Setup

As shown in Fig. 1, CReS-VCT is designed as a hierarchical HIL testbed for modeling large-scale power systems. CReS-VCT co-simulates DER/distribution/transmission networks, physical inverters, controllers, and communication links. The main function of CReS-VCT is to coordinates control actions of Volt-VAR devices and DERs from the sub-transmission system down to DERs located at distribution feeders.

An IEEE 118-bus HIL test system was designed by PNNL at Richland, WA to simulate the sub-transmission system. A substation HIL test system was developed by NC State at Raleigh, NC to simulate three distribution feeders in detail using three IEEE 123-node systems. A 60-Hz AC source is placed at the feeder head to represent the transmission system such that the voltage magnitude and phase angle captured from the transmission system can be passed on to the distribution feeder. The total consumption of the three feeders will be sent back to the transmission system every 100ms, the time of which is determined by the Modbus pulling frequency. A single-phase inverter at UT-Austin is assumed to be connected to one of the feeder systems in the NC State distribution HIL system via an Ethernet-based link. The photovoltaic (PV) inverter hardware represents one phase of a three-phase PV inverter model in the NC State distribution HIL system.

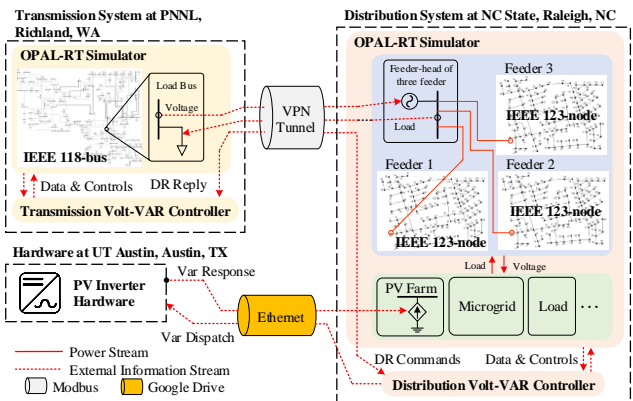

Fig. 1. Architecture overview for CReS-VCT.

### B. External Controller Setup

Control algorithms are implemented externally to the HIL testbeds, therefore, power-flow solvers such as OpenDSS and optimization solvers such as GAMS and Knitro can be used to develop and implement those algorithms.

At the transmission system level, a Volt-VAR controller communicates with both the transmission HIL test system and the distribution VVC controller so it can optimally dispatch the system-wide resources at both the transmission and distribution levels to meet transmission-level voltage control requirement, minimize voltage fluctuations, and minimize losses with minimum cost [4]. After solving the VVC problem at the transmission level, the transmission VVC controller sends demand response (DR) commands (e.g., request for reactive power and active power curtailment) to the distribution controller. The distribution controller tries to meet the sub-transmission power control request, minimize nodal voltage fluctuations, and minimize system losses with minimum control cost [5]. The distribution VVC controller reads measurements from the HIL simulator, calculates the voltage-sensitive matrix (VSM) using OpenDSS, executes a mixed-integer nonlinear program, and sends control commands to the distribution HIL system and the PV inverter hardware. At each control interval (i.e., 5min), this control sequence repeats.

### C. Information Flows and Communication Links

Under the CReS-VCT architecture, two external controllers communicate with the corresponding HIL test systems either locally through Ethernet connections or remotely via a VPN tunnel. Information flows, including both monitoring and control signals, are shown in Fig. 2. Modbus is used as the communication protocol for passing messages between the transmission and distribution VVC controllers and between the distribution controllers and the controllable devices modeled in distribution HIL test system.

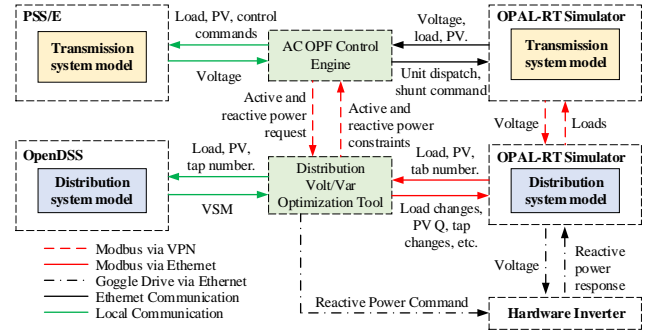

Fig. 2. Control diagram for CReS-VCT.

To integrate the test systems located at the different partner sites, we developed two methods: file-based and VPN-based. File-based links can be set up via Google Drive. The communication interval between two test systems can be varied due to the communication delays caused by the shared file updated rate. The establishment of a VPN-based connection requires coordination among information technology departments of different organizations so there is an associated cost. The communication delay of this method is minimized (typically in the range of milliseconds).

In the CReS-VCT setup, the transmission and distribution HIL test systems are connected using the VPN-based link so communication delays and their impact on co-simulation results of transmission and distribution grids is minimal. The file-based link is used between the PV inverter and the distribution HIL test system so that the setup is simple and flexible.

### III. Modeling Methodology

This section introduces the modeling methodology of the transmission, distribution, and DER systems.

#### A. Transmission and Distribution Systems Modeling

The OPAL-RT ePHASORsim is a model-based time-series simulation tool that can conduct phasor-based unbalanced power flow calculations and dynamic simulation. Thus, it is used to simulate the transmission and distribution systems. The advantage of using ePHASORsim is that it can link different source files developed in other software packages (e.g., PSS/E and CYME) and execute the integrated model in a real-time simulation environment.

As shown in Fig. 1, an IEEE 118-bus test system is used to model the transmission network. The model is first developed in PSS\E and then imported to the transmission HIL simulator. To enable PV integration-related studies, PV generators are added to 54 load buses. Details of selecting solar generator locations and capacities for modeling different solar penetration scenarios can be found in [6].

When using ePHASORsim to simulate a transmission system, if a generator is equipped with a governor, the transmission controller cannot dispatch the generator power. Therefore, the generator will be split into two generators; one generator has the governor system and the other one does not. Detailed settings for generator are available in [7]. The original IEEE 118 bus test system has only one snapshot power flow data. To conduct time-series analysis, yearly data at 5-minute resolution time are generated from utility data sets for both loads and solar farms using methods introduced in [6].

As shown in Fig. 1, to simulate distribution grids in more detail, we use three IEEE standardized 123-node feeders to model the distribution systems connected to Bus 84 in the IEEE 118 Bus transmission system. Moreover, we modeled 85 load nodes (80 single-phase load nodes and five three-phase load nodes) in each of the three distribution feeders. The feeder models are first developed in CYME and then imported to the distribution HIL simulator. To model the solar farms connected to the distribution feeders, we put three solar farms onto three three-phase nodes (e.g., Nodes 44, 54, 67). Note that there is no load connected to those nodes. Solar generators are modeled as negative constant power loads using the default CYME model.

To better model load diversity in the distribution feeders, a feeder load disaggregation algorithm is used to disaggregate the feeder head load profile to each load node using smart meter data provided by Duke Energy. Details for the disaggregation method were introduced in [8].

#### B. Modeling of PV Inverter

PV systems are simulated in OPAL-RT eMEGAsim. IEEE-1547 has recently been revised to require inverters to include VVC functionality [9]. Thus, all PV inverters in the proposed test system are designed so that they can provide VVC services.

A modified PV power constraint model is used to maximize the use of the PV inverter with any power factor for VVC applications [10]. Then, we now have

$$\left(\frac{P^{PV}}{S^{PV}}\right)^2 + \left(\frac{Q^{PV}}{k \cdot S^{PV}}\right)^2 \leq 1 \tag{1}$$

where $k$ is the improvement factor for reactive power and is set at 1.1 for a normal insulated-gate bipolar transistor (IGBT) based two-Level inverter [10]. As shown in Fig. 3, the operating area for the modified inverter model is a semi-ellipse instead of the semi-circle that a traditional PV inverter would follow.

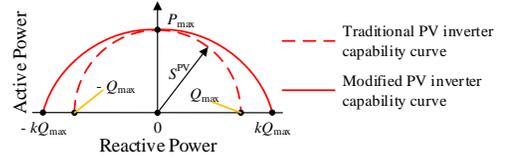

Fig. 3. PV inverter active and reactive power constraint model.

Therefore, when a PV inverter operates at normal conditions, it produces $P^{PV}$ and zero reactive power, and the inverter adjusts its generation dynamically as irradiance changes within each time period. If the PV inverter receives a reactive power command $Q_{Cm}^{PV}$ from the distribution VVC controller, the actual active power $P_{Ac}^{PV}$ and reactive power $Q_{Ac}^{PV}$ can be calculated as

$$\begin{aligned} Q_{Ac}^{PV} &= \max(\min(Q_{Cm}^{PV}, k \cdot S^{PV}), -k \cdot S^{PV}) \\ P_{Ac}^{PV} &= \min\left(\sqrt{\left(S^{PV}\right)^2 - \left(\frac{1}{k} Q_{Ac}^{PV}\right)^2}, P^{PV}\right) \end{aligned} \tag{2}$$

#### C. Modeling of Power System Loads

Distribution spot loads are modeled using the default CYME model and each distribution feeder has 57 constant power loads, 13 constant current loads, and 15 constant impedance loads. We assigned the rated consumption power to each spot load every 5 minutes using nodal load profiles; thus, for constant current loads and constant impedance loads, load consumption may change during every interval with the change of nodal voltage.

Controllable loads are simulated in OPAL-RT eMEGAsim. A microgrid is assumed to connect to Node 49, and is modeled as a constant power load. Details are available in [3], [11].

#### D. Systems Simulation and Control Coordination

The networked HIL testbed enables a closed-loop simulation for large-scale power systems in which transmission, distribution, and DER systems are co-simulated. Because different modeling approaches are applied (i.e., phasor-based and electromagnetic simulations), each system has its modeling setup with modeling time steps ranging from microseconds to milliseconds. For example, transmission and distribution systems are modeled using ePHASORsim with a time step of 10ms, while DERs are modeled using eMEGAsim with a time step of 50μs. A detailed description of the temporal couplings is described in [3].

As shown in Fig. 4, at the beginning of a simulation cycle, the distribution controllers pull measurements (e.g., loads, PVs and voltage) from distribution HIL systems to start a computing cycle for calculating the active and reactive power (P&Q) constraints. Also, the transmission controller sends generator commands to the transmission HIL system. After the distribution controller computing cycle ends, the obtained P&Q constraints are sent to the transmission controller so an optimal VVC problem can be solved. Once the VVC solution is obtained, the transmission-level VVC commands are sent to the transmission HIL system for execution and P&Q requests are sent to the distribution controller to be executed. Once the distribution controller receives the P&Q request, it solves a VVC problem such that DR, PV, and the shunt commands can be generated and sent to the distribution HIL system [4].

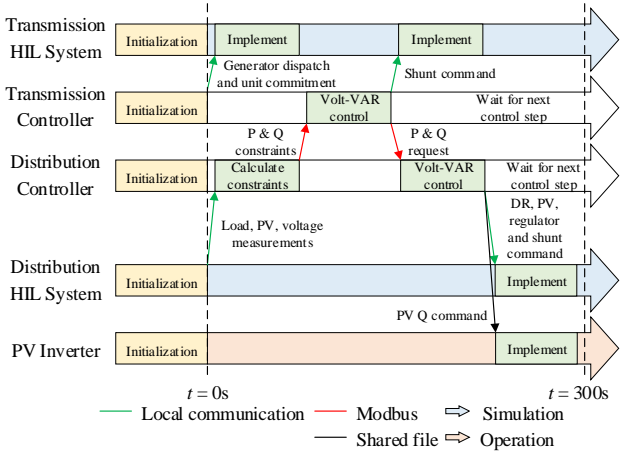

Fig. 4. Control coordination timeline for HIL simulators and inverter.

The control sequence and coordination with PV inverter hardware implementations are shown in Fig. 5. After solving a VVC problem, the distribution controller sends the reactive power command to the PV inverter through a file-based link.

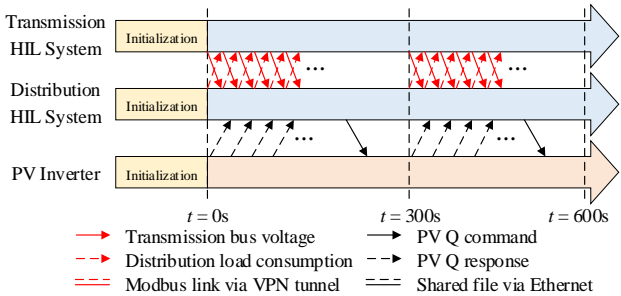

Fig. 5. Co-simulation coordination timeline for HIL simulators and inverter.

The information flows communicated between transmission and distribution system are shown in Table I, and between the distribution and the PV hardware implementation are shown in Table II. Note that the update time steps are the intervals that controllers and HIL systems update their simulation results and control commands for communication proposes, and those time steps are chosen as our target to implement CReS-VCT in real-time with 5-minute control intervals. For a networked HIL test system, depending on the monitoring and control sequences of each control system, the update time for each subsystem can be modified and the communication interval adjusted accordingly. However, some communication delays are limited by network conditions. For example, the average communication latency is 110ms when using a VPN tunnel between PNNL and NC State, while the average communication delay is about 1 minute when using Google Drive to communication between NC State and UT-Austin.

TABLE I. COMMUNICATION BETWEEN TRANSMISSION AND DISTRIBUTION

| Communication Direction | Single | Unit | Update Time |
|---|---|---|---|
| From transmission system to distribution system | System simulation time | second | 100ms |
| | Sub-transmission bus voltage | p.u. | |
| | Voltage angle for Phase $a$ | ° | |
| | Transmission scenario counter | / | 5min |
| | Active power curtailment request | p.u. | |
| | Reactive power request | p.u. | |
| From distribution system to transmission system | System simulation time | second | 100ms |
| | Total active power for feeders | kW | |
| | Total reactive power for feeders | kVAR | |
| | Distribution scenario counter | / | 5min |
| | PV $P$ curtailment upper limit | kW | |
| | PV $Q$ upper and lower limit | kVAR | |
| | DR $P$ upper and lower limit | kW | |
| | Total losses for feeders | kW | |

TABLE II. COMMUNICATION BETWEEN DISTRIBUTION AND PV HARDWARE

| Communication Direction | Single | Unit | Update Time |
|---|---|---|---|
| From distribution system to PV inverter | System simulation time | second | 5min |
| | PV $Q$ request ($Q_{Cm}^{PV}$) | p.u. | |
| From PV inverter to distribution system | Inverter execution time | second | 1min |
| | PV $Q$ response ($Q_{Re}^{PV}$) | p.u. | |

## IV. CASE STUDIES AND ANALYSIS

We used a 48-hour data set (April 6-7) to conduct a test case for the proposed CReS-VCT. At transmission Bus 84, solar power generated could exceed the load at noon; thus, the distribution feeders have reverse power flow to the transmission grid. Aggregated profiles at the feeder head for load and solar generation are shown in Fig. 6. Load profiles for each feeder spot load are designed based on the feeder head load profile. The median of nodal load profiles within three feeders and the range between the first and third quartile are depicted in Fig. 7.

Fig. 8 shows the feeder reactive power at each control step. A positive value means the distribution systems are absorbing reactive power and the negative means generating. The red curve represents the original no-VVC reactive load baseline for all three feeders. When the transmission controller sends a request to the distribution controller (the magenta circle in Fig. 8) [4], the distribution controller manages DERs to fulfill the request (the brown line in Fig. 8).

The error between the reactive power response and request (the black line in Fig. 8) is near zero most of the time except the period between $t \in [35, 40]$, during which, the communication link (the black dash-dot lines in Fig. 2) between the distribution HIL system and the inverter hardware is severed. Therefore, the distribution controller cannot dispatch the hardware and the distribution HIL cannot receive the reactive power response from the inverter hardware. The maximum reactive power deficit is roughly 600kVAR. The actual total feeder reactive power (the cyan line in Fig. 8) is the sum of the reactive load baseline and the PV inverter reactive power response.

The nodal voltages for three feeders at one-minute resolution are shown in Fig. 9. As shown in Fig. 9, the distribution controller can control nodal voltages within the allowed limits. However, for Feeder 1, to which the PV inverter hardware implementation is connected, the nodal voltages in Feeder 1 have more spikes compared to Feeders 2 and 3 because of the updating delay for inverter reactive power response. Moreover, nodal voltages in Feeder 1 violate the allowed upper limit (e.g., 1.05 p.u.) when the file-based link is interrupted.

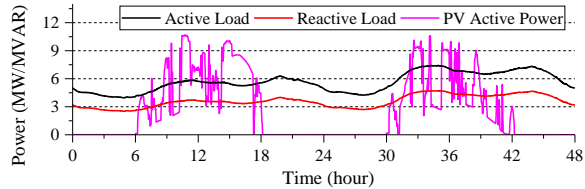

Fig. 6. Load profile and solar generation on transmission Bus 84.

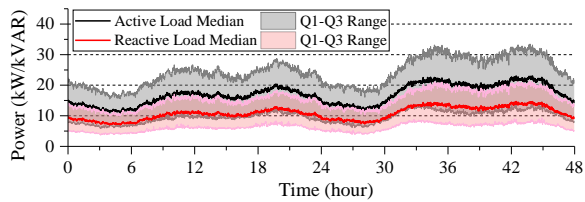

Fig. 7. Median of nodal load profile and the range of the first and third quartile.

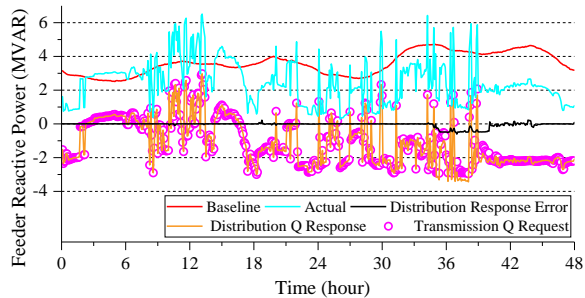

Fig. 8. Distribution feeder reactive power.

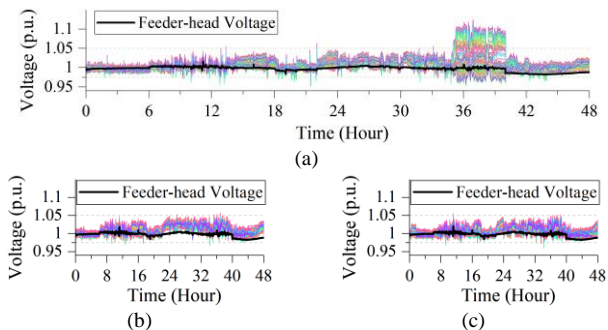

Fig. 9. Nodal voltages for (a) Feeder 1, (b) Feeder 2, and (c) Feeder 3.

## V. CONCLUSION

In this paper, we present a networked HIL test system that uses different test systems developed by three different teams and multiple OPAL-RT-based HIL simulators located in three different locations. Overall, the performance of the networked HIL simulation system meets our simulation requirements. The presence of asynchronous computing time for different control algorithms, communication latency and losses, and uncertainty in information exchange cycles and update rates are present in real-world implementations. We were able to test different control design mechanisms to avoid deterioration in controller performance. Instead of developing the whole test system solely at NC State or at PNNL, we collaborated with each other to significantly shorten the time required to develop and implement the test system.

Because of page limitations, we could not include detailed discussions on mitigation methods for interrupting communication. In a follow-up paper, we will discuss how to design and develop a robust, networked HIL simulation test system that models thousands of buses at both transmission and distribution systems with hundreds of power electronics systems connected to the grid at different levels.

## ACKNOWLEDGMENT

The authors would like to thank Pacific Northwest National Laboratory, University of Texan at Austin Semiconductor Power Electronics Center, Total S.A., and the Department of Energy SunLamp Program for their support to this work.